\documentclass[aip,jmp]{revtex4-1}
\usepackage{graphicx}
\usepackage{amsmath}
\usepackage{amsfonts}
\usepackage{amssymb}
\usepackage{MnSymbol}

\newcommand{\be}{\begin{equation}}
\newcommand{\ee}{\end{equation}}

\begin{document}

\title{Statistics of time delay and scattering correlation functions in chaotic systems I. Random Matrix Theory}
\author{Marcel Novaes}
\affiliation{Instituto de F\'isica, Universidade Federal de Uberl\^andia\\ Av. Jo\~ao
Naves de \'Avila 2121, Uberl\^andia, MG, 38408-100, Brazil}
\begin{abstract}
We consider the statistics of time delay in a chaotic cavity having $M$ open channels, in the absence of time-reversal invariance. 
In the random matrix theory approach, we compute the average value of polynomial functions of the time delay matrix $Q=-i\hbar S^\dag dS/dE$, where $S$ is the scattering matrix. Our results do not assume $M$ to be large. In a companion paper, we develop a semiclassical approximation to $S$-matrix correlation functions, from which the statistics of $Q$ can also be derived. Together, these papers contribute to establishing the conjectured equivalence between the random matrix and the semiclassical approaches.
\end{abstract}
\maketitle

\section{Introduction}

Quantum scattering processes at energy $E$ can be described by the scattering matrix
$S(E)$, which transforms incoming wavefunctions into outgoing wavefunctions. This matrix
is necessarily unitary, in order to enforce conservation of probability and,
consequently, conservation of charge. Another important operator is the Wigner-Smith time
delay matrix $Q$, \cite{wigner,smith} a hermitian matrix related to the energy derivative of
$S$ as $Q=-i\hbar S^\dag dS/dE$. Its eigenvalues are the delay times of the system, and its normalized trace is the Wigner time
delay, $\tau_W=\frac{1}{M}{\rm Tr}Q$. These quantities contain information about the time a particle spends
inside a scattering region. A thorough discussion can be found in the review
\cite{review2}.

We consider a scattering region (`cavity') inside of which the classical dynamics is
strongly chaotic, connected to the outside world by small, perfectly transparent, openings.  This can be realized
in experiments with microwave cavities \cite{cavities1,cavities2,cavities3,cavities4},
quantum dots \cite{dots1,dots2,dots3,dots4} and compound nuclei \cite{nuclear}. We assume $M$ open channels, so that $S$ and $Q$ are $M$-dimensional.
We also assume there is a well defined classical decay rate $\Gamma$, such that the total
probability of a particle to be found inside the cavity decays exponentially in time as
$e^{-\Gamma t}$. The quantity $\tau_D=1/\Gamma$ is called the classical `dwell
time'.

When the wavelength is much smaller than the cavity size, the $S$ and $Q$ matrices are strongly oscillating
functions of the energy, and a statistical approach is advantageous. One such approach is
based on random matrix theory (RMT). Its main hypothesis is that $S$ behaves like a
random unitary matrix, distributed in the unitary group $\mathcal{U}(M)$ according to some probability
measure (in the presence of time-reversal invariance, $S$ must also be symmetric; we do
not consider that situation in this work). For ideal openings,
this distribution is the normalized Haar measure of the group. We denote the 
averages with respect to this distribution by the symbol $\langle \cdot\rangle$.
In particular, the average of the
Wigner time delay is equal to the classical dwell time \cite{caio1,caio2}, $\langle
\tau_W\rangle=\tau_D.$

This RMT approach has had much success in describing transport statistics
\cite{beenakker,prb78mn2008,prb80bak2009,macedo}, such as conductance, shot-noise, etc. 
It can also be applied to time delay statistics, but usually not starting
from the $S$ matrix, but rather from the Hamiltonian of the system. This allows better
control of the energy dependence and calculation of correlation functions, but requires
mapping the problem to a nonlinear supersymmetric $\sigma$-model
\cite{sigma1,sigma2,sigma3,sigma4}.

On the other hand, Brouwer, Frahm and Beenakker \cite{brouwer} have found the
joint probability distribution for the eigenvalues of $Q$, let us
denote them by $\tau_1,...,\tau_M$.
This allowed the calculation of marginal distributions \cite{brouwer2}, distribution of
Wigner time delay  (for $M=2$, \cite{Mis2} and in the limit $M\gg 1$, \cite{majumdar}) and the average value of moments, \cite{simm1,simm2} \be \mathcal{M}_n=\frac{1}{M}{\rm
Tr}[Q^n]=\sum_{i=1}^M \tau_i^n.\ee A few other polynomial functions have also been
computed \cite{simm3,garcia}. A recent review, also considering extension to non-ideal openings and other symmetry classes, can be found in \cite{dima}.

We hereby advance the RMT approach to statistics of time delay,
obtaining an explicit formula for the average value of general polynomial quantities of the
kind \be\label{moments} \mathcal{M}_{n_1,n_2,...}=\frac{1}{M}{\rm Tr}[Q^{n_1}]\frac{1}{M}{\rm
Tr}[Q^{n_2}]\cdots,\ee for any finite set of positive integers $n_1,n_2,...$ Our
method starts from the result of \cite{brouwer} and is based on Schur function expansions
and determinant evaluations. Importantly, our results are not perturbative in the number
of channels, being valid at finite values of $M$.

In the next Section we briefly present our
results. Section 3 contains some preliminary material, and in Section 4 we present our calculations. 

\section{Results}

The average value of the moments $\mathcal{M}_n$ have been found 
for general number of channels $M$,\cite{simm1} but expressed as a sum with $M$ terms. Our results imply
the following simple general formula, which contains a more efficient sum, with only $n$ terms:
\be\label{ourmoments} \langle\mathcal{M}_n\rangle=\tau_D^n\frac{M^{n-1}}{n!}\sum_{k=0}^{n-1}(-1)^k{n-1
\choose k}\frac{[M-k]^n}{[M+k]_n},\ee where \be [x]^n=x(x+1)\cdots(x+n-1), \quad
[x]_n=x(x-1)\cdots(x-n+1),\ee are the raising and falling factorials.

The first four cumulants of the Wigner time delay have been computed
using some nonlinear differential equation for their generating function.\cite{simm3}
This amounts to finding the value of $\langle \tau_W^j\rangle$ for $j$ up to 4. Our
results imply the explicit general formula \be \label{wigmom}\langle
\tau_W^n\rangle=\frac{\tau_D^n}{n!}\sum_{\lambda\vdash
n}d_\lambda^2\frac{[M]^{\lambda}}{[M]_{\lambda}},\ee where the sum is over all partitions
of $n$, the length of a partition $\lambda$ is denoted $\ell(\lambda)$ and \be\label{factorials}
[M]^\lambda=\prod_{i=1}^{\ell(\lambda)}[M-i+1]^{\lambda_i},\quad
[M]_\lambda=\prod_{i=1}^{\ell(\lambda)}[M+i-1]_{\lambda_i}\ee are generalizations of the
rising and falling factorials. The quantity $d_\lambda$ is the dimension of the
irreducible representation of the permutation group labeled by $\lambda$, and it is given
by \be\label{dimension}
d_\lambda=n!\prod_{i=1}^{\ell(\lambda)}\frac{1}{(\lambda_i-i+\ell(\lambda))!}\prod_{j=i+1}^{\ell(\lambda)}
(\lambda_i-\lambda_j-i+j).\ee

The above examples are derived from particular cases of our most general result, which is
the average value of a general Schur function of $Q$:\be\label{RMTSchur}  \langle
s_\lambda(Q)\rangle=(M\tau_D)^n\frac{d_\lambda}{n!}\frac{[M]^{\lambda}}{[M]_{\lambda}}.\ee
These functions are homogeneous symmetric polynomials in the eigenvalues of $Q$. Since any symmetric polynomial in these variables
can be expressed as a linear combination of Schur functions, this can be seen as a
complete solution to the problem of computing the average value of polynomials (or analytic functions, if we allow infinite series) in
$Q$, such as the quantities $\mathcal{M}_{n_1,n_2,...}$ defined in (\ref{moments}). For instance, the first of these which are neither of the form (\ref{ourmoments}) nor of the form (\ref{wigmom}) are \be\label{M21} \langle \mathcal{M}_{2,1}\rangle=\frac{2M^2(M^2+2)}{(M^2-1)(M^2-4)},\ee and \be\label{M22} \langle \mathcal{M}_{2,2}\rangle=\frac{4M^2(M^4+8M^2-3)}{(M^2-1)(M^2-4)(M^2-9)},\quad \langle \mathcal{M}_{3,1}\rangle=\frac{6M^2(M^2+1)^2}{(M^2-1)(M^2-4)(M^2-9)}.\ee

\section{Preliminaries}
\subsection{Partitions and permutations}

A weakly decreasing sequence of positive integers, $\lambda = (\lambda_1, \lambda_2,...)$
is called a partition of $n$, denoted by $\lambda\vdash n$ or by $|\lambda|=n$, if $\sum_i\lambda_i=n$. Each of
the integers is a part, and the total number of parts is the length $\ell(\lambda)$.

Partitions of $n$ label the conjugacy classes of the permutation group $S_n$: the cycle
type of a permutation $\pi$ is a partition whose parts are the lengths of the cycles of
$\pi$, and two permutations $\pi,\sigma$ have the same cycle type if and only if they are
conjugated, i.e. if there exists $\tau$ such that $\pi=\tau\sigma\tau^{-1}$. Let
$\mathcal{C}_\lambda$ denote the set of permutations with cycle type $\lambda$, and
$|\mathcal{C}_\lambda|$ the number of elements in $\mathcal{C}_\lambda$.

For any finite group, there are as many irreducible representations as there are
conjugacy classes. Therefore, partitions of $n$ also label the irreducible
representations of $S_n$. The trace of permutation $\pi$, in the representation labeled
by $\lambda$, is denoted as $\chi_\lambda(\pi)$ and called its character. The character
of the identity, $\chi_\lambda(1)=d_\lambda$, is the dimension of the representation, for
which there is the explicit formula (\ref{dimension}). Characters of $S_n$ are class
functions, i.e. $\chi_\lambda(\pi)$ depends only on the cycle type of $\pi$ and we may
write $\chi_\lambda(\mu)$ if $\pi\in\mathcal{C}_\mu$. Characters satisfy
orthogonality relations, \be \sum_{\tau\in S_n}\chi_\mu(\tau) \chi_\lambda(\tau\sigma)=
\frac{n!}{d_\lambda}\chi_\lambda(\sigma)\delta_{\mu,\lambda}.\ee

\subsection{Symmetric functions}

Let $X$ be a matrix of dimension $N$, with eigenvalues $x_i$, $1\leq i\leq N$. Power sum
symmetric functions of matrix argument are defined as \be
p_\lambda(X)=\prod_{i=1}^{\ell(\lambda)} p_{\lambda_i}(X),\quad  p_n(X)={\rm
Tr}[X^n]=\sum_{i=1}^Nx_i^n.\ee They are clearly symmetric functions of the eigenvalues.

Another important family of symmetric functions are Schur functions, related to power
sums by \cite{Sagan} \be\label{power2schur} s_\lambda(X)=\frac{1}{n!}\sum_{\mu\vdash
n}|\mathcal{C}_\mu|\chi_\lambda(\mu)p_\mu(X), \quad p_\lambda(X)=\sum_{\mu\vdash
n}\chi_\mu(\lambda)s_\mu(X).\ee These functions can also be written as a ratio of
determinants, \be\label{schurasdet}
s_\lambda(X)=\frac{\det(x_i^{\lambda_j-j+N})}{\Delta(X)},\ee where \be
\Delta(X)=\det(x_i^{j-1})=\prod_{i=1}^N\prod_{j=i+1}^N(x_j-x_i),\ee is the Vandermonde
determinant. 

The value of the Schur function when all arguments are equal to $1$ is
\be\label{s1N} s_\lambda(1^N)=\frac{d_\lambda}{n!}[N]^\lambda,\ee where $[N]^\lambda$ is
the generalization of the rising factorial defined in (\ref{factorials}). Noticing that
in the formula for $d_\lambda$ there appears the Vandermonde for $x_i=\lambda_i-i$, it is
also possible to show that \be \label{special}
\Delta(\{\lambda_i-i\})=s_\lambda(1^N)\prod_{j=1}^{N-1}j!.\ee

Let $d\vec{x}=dx_1\cdots dx_N$. In view of the identity \be\label{andreief} \int
d\vec{x}\det(f_i(x_j))\det(g_i(x_j)) = N!\det\left( \int dx f_i(x)g_j(x)\right),\ee
easily proved using the Leibniz formula for the determinant, the representation
(\ref{schurasdet}) of Schur functions is useful for performing multidimensional
integrals involving these functions.

\section{Statistics of the time delay matrix}

\subsection{Average of Schur functions}

We wish to compute the average value of a Schur function of the time delay matrix,
$s_\lambda(Q)$. This will be done using the following result obtained in \cite{brouwer}: the 
probability distribution of the matrix $\gamma=Q^{-1}$ is \be
P(\gamma)=\frac{1}{\mathcal{Z}}|\Delta(\gamma)|^2\det(\gamma)^Me^{-M\tau_D{\rm Tr}\gamma}.\ee
where \be \mathcal{Z}=\int_0^\infty |\Delta(\gamma)|^2\det(\gamma)^Me^{-M\tau_D{\rm
Tr}\gamma}d\gamma\ee is a normalization constant.

Let $\tau_i$, $1\leq i\leq M$, be the eigenvalues of $Q$ and $\gamma_i=1/\tau_i$ be the
eigenvalues of $\gamma$. The normalization constant is computed using
Eq.(\ref{andreief}):\be\mathcal{Z}=\int_0^\infty
d\vec{\gamma}\det(\gamma_j^{M+i-1}e^{-M\tau_D\gamma_j})\det(\gamma_i^{j-1})=
\frac{M!}{(M\tau_D)^{2M^2}}\det((M+j+i-2)!).\ee Standard determinant manipulations yield \be
\mathcal{Z}=\frac{1}{(M\tau_D)^{2M^2}}\prod_{j=1}^M j!(M+j-1)!.\ee

The quantity we are after is \be \langle
s_\lambda(Q)\rangle=\frac{1}{\mathcal{Z}}\int_0^\infty d\vec{\gamma}
|\Delta(\gamma)|^2\det(\gamma)^Me^{-M\tau_D{\rm Tr}\gamma}s_\lambda(\gamma^{-1}).\ee
Writing the Schur function as a determinant, as in Eq.(\ref{schurasdet}), and using the
following identity for the Vandermonde, \be
\Delta\left(\gamma^{-1}\right)=\frac{(-1)^{M(M-1)/2}\Delta(\gamma)}{\det\gamma^{M-1}},\ee
we arrive at \be \langle
s_\lambda(Q)\rangle=\frac{(-1)^{M(M-1)/2}}{\mathcal{Z}}\int_0^\infty d\vec{\gamma}
\det(\gamma_j^{2M+i-2}e^{-M\tau_D\gamma_j})\det(\gamma_i^{-\lambda_j+j-M}).\ee
Using Eq.(\ref{andreief}) again we have \be\label{intermediate} \langle
s_\lambda(Q)\rangle=\frac{(-1)^{M(M-1)/2}}{\mathcal{Z}(M\tau_D)^{2M^2-n}}M!\det((M-\lambda_j+j+i-2)!).\ee

Consider the determinant $\det((x_j+i)!)$. Suppose we factor out a term $(x_j+1)!$ from
each row. The remaining determinant has the following structure: its $ij$ element is a
monic polynomial in $x_j$ of degree $i-1$. It is well known \cite{Mehta} that it therefore must be
equal to the Vandermonde $\Delta(x)$. 

Applying the above argument to (\ref{intermediate}), we
get \be \langle
s_\lambda(Q)\rangle=\frac{1}{\mathcal{Z}(M\tau_D)^{2M^2-n}}s_\lambda(1^M)\prod_{j=1}^Mj!(M-\lambda_j+j-1)!,\ee
where we used $ \Delta(\{M-\lambda_i+i-2\})=(-1)^{M(M-1)/2}\Delta(\{\lambda_i-i\})$ and the
special value of the Vandermonde, Eq. (\ref{special}). Plugging in the values of
$s_\lambda(1^M)$ and $\mathcal{Z}$, we arrive at \be \langle
s_\lambda(Q)\rangle=(M\tau_D)^n\frac{d_\lambda}{n!}[M]^\lambda\prod_{j=1}^M
\frac{(M-\lambda_j+j-1)!}{(M+j-1)!},\ee or, in terms of the generalized falling factorial
defined in (\ref{factorials}), our claimed result, \be\langle
s_\lambda(Q)\rangle=(M\tau_D)^n\frac{d_\lambda}{n!}\frac{[M]^\lambda}{[M]_\lambda}.\ee

\subsection{Particular cases}

The relation between power sums and Schur functions, Eq. (\ref{power2schur}), allows the
calculation of more familiar quantities, such as \be\label{Mn} \langle
\mathcal{M}_n\rangle=\frac{1}{M}\langle p_{n}(Q)\rangle=\frac{1}{M}\sum_{\lambda\vdash n}\chi_\lambda(n)\langle
s_\lambda(Q)\rangle.\ee The character $\chi_\lambda(n)$ is different from zero only if
$\lambda=(n-k,1^k)$ (so-called hook partitions), and is equal to $(-1)^k$ in this case.
On the other hand, the dimension $d_\lambda$ becomes ${n-1 \choose k}$ for hooks, and
with this we arrive at our example (\ref{ourmoments}). 

The other example we mentioned in Section 2 was
\be \langle \tau_W^n\rangle=\frac{1}{M^n}\langle p_{(1,1,...,1)}(Q)\rangle=\frac{1}{M^n}\sum_{\lambda\vdash
n}d_\lambda\langle s_\lambda(Q)\rangle.\ee Finally, consider the general moments $\mathcal{M}_{n_1,n_2,...}$. We may assume that $\mu=(n_1,n_2,...)$ is a partition of some integer, $|\mu|$. Then, we have $\mathcal{M}_{n_1,n_2,...}=\frac{1}{M^{\ell(\mu)}}p_\mu(Q)$, and \be \langle \mathcal{M}_{n_1,n_2,...}\rangle=\frac{1}{M^{\ell(\mu)}}\sum_{\lambda\vdash|\mu|}\chi_\lambda(\mu)\langle s_\lambda(Q)\rangle.\ee Using this expression, we recover our examples (\ref{M21}) and (\ref{M22}).

\section*{Acknowledgments}

Financial support from Conselho Nacional de Desenvolvimento Cient\'ifico e Tecnol\'ogico (CNPq) is gratefully acknowledged.

\end{document}